\documentclass[12pt,a4plain]{article}
\usepackage{graphicx} 
\usepackage{setspace}

\usepackage[top=4cm, bottom=1.5cm, left=2.5cm, right=2.5cm]{geometry} 
 
\begin{document}

\title{\bf First principles study of Si(335)-Au surface}

\author{
\vspace{0.5cm} 
{\bf Mariusz Krawiec\footnote{Electronic address: 
        krawiec@kft.umcs.lublin.pl}} \\
        Institute of Physics and Nanotechnology Center, \\
        M. Curie-Sk{\l}odowska University, Pl. M. Curie-Sk{\l}odowskiej 1, \\
	20-031 Lublin, Poland}

\maketitle

\begin{abstract}
The structural and electronic properties of gold decorated Si(335) surface are 
studied by means of density-functional calculations. The resulting structural 
model indicates that the Au atoms substitute some of the Si atoms in the middle 
of the terrace in the surface layer. Calculated electronic band structure near the 
Fermi energy features two metallic bands, one coming from the step edge Si 
atoms and the other one having its origin in hybridization between the Au and 
neighboring Si atoms in the middle of the terrace. The obtained electronic bands 
remain in good agreement with photoemission data. 
\end{abstract}


\newpage


\section{\label{intro} Introduction}

Physics of one-dimensional (1D) objects is predicted to be qualitatively 
different form that of higher dimensions \cite{Giamarchi}. The most spectacular 
examples include a breakdown of the Fermi liquid theory
\cite{Luttinger,Auslaender} and Peierls metal-insulator transition
\cite{Peierls}. Theorists found the one-dimensional physics to be very elegant, 
as many problems can be solved in 1D but not in higher dimensions. On the other
hand, experimentalists face with great difficulties in fabricating truly
one-dimensional structures. The most promising way is a creation of such
structures on semiconducting or insulating substrates \cite{Himpsel,Owen}. In 
this case, the chain atomic structure is governed by the substrate lattice. More 
importantly, the electrons near the Fermi energy are completely decoupled
from the substrate, as there is a band gap in the electronic spectrum of the
substrate. Of course, there are low energy states responsible for holding chain
atoms on the surface. Fortunately, only the states near the Fermi level 
determine electronic properties of the system. 

Recently, the most extensively studied templates for fabrication of one-dimensional 
objects are vicinal Si(111) surfaces \cite{Crain_1,Leroy}. Those include gold 
induced chains on Si(111) \cite{Altmann,Erwin}, Si(335) 
\cite{Zdyb}-\cite{Kisiel_1}, Si(553) \cite{Crain_2}, 
\cite{Crain_3}-\cite{Crain_4}, Si(557) \cite{Jalochowski}-\cite{Sauter}, 
Si(775) \cite{Crain_1,Crain_3} or Si(5512) \cite{Baski}-\cite{Jeong}. 

The stepped Si(335)-Au reconstruction is observed at the gold coverage of 0.28 ML 
and consists of Si(111) terraces $3 \frac{2}{3} \times a_{[1 1 \bar{2}]}$ 
(1.26 nm) wide. Each terrace contains a single row of gold atoms running 
parallel to the step edge, i.e. in the $[1 \bar{1} 0]$ direction. A number of 
techniques has been employed to investigate structural and electronic 
properties of Si(335)-Au surface. Those include high-energy electron 
diffraction (RHEED) \cite{Zdyb}, angle-resolved photoemission spectroscopy 
(ARPES) \cite{Crain_2,Kisiel_1} and scanning tunneling microscopy (STM) 
\cite{Crain_2,MK_1}. In particular, STM topography data show a single 
monatomic chain on each terrace, which has been interpreted as the Si atoms
with broken bonds at the step edge \cite{Crain_2}. As the Au atoms are more
electronegative than Si, they pair neighboring Si bond with their $6s$, $p$ 
electrons. As a result, a low-lying bound state, not visible to STM, is 
created. The STM experiments show a single chain within each terrace, so one
would naively expect a single metallic band in electronic structure. However, 
ARPES data show not one but two bands crossing the Fermi level \cite{Crain_2}. 
It is the purpose of the present work to solve this puzzle and to identify the 
bands observed in the ARPES experiment.

While the surfaces like Si(553)-Au or Si(557)-Au were extensively studied 
theoretically \cite{Crain_2,Riikonen,Crain_4,Portal_1,Portal_2,Portal_3}, there 
are no first principles investigations of the Si(335)-Au surface. Moreover, 
there is no structural model confirmed by first principles calculations.
However, the model of this surface, based on an analogy with Si(557)-Au surface, 
has been proposed in Ref. \cite{Crain_2}. It is simply a truncation of the 
Si(557)-Au structure. So the second purpose of the present work is to check 
whether the model of Ref. \cite{Crain_2} is a good candidate for the atomic 
reconstruction of Si(335)-Au surface. In the following, I will focus on the 
presentation of the results regarding structural model and electronic 
properties of the system.


\section{\label{method} Details of calculations}

The calculations have been performed using the SIESTA code 
\cite{Ordejon}-\cite{Soler}, which performs standard pseudopotential density 
functional calculations using a linear combination of numerical atomic orbitals
as a basis set. I have used here the generalized gradient approximation (GGA) 
to DFT \cite{Kohn,Perdew}, Troullier-Martins norm-conserving pseudopotentials
\cite{Troullier}, and a double-$\zeta$ polarized (DZP) basis set for all the
atomic species \cite{Portal_4,Artacho,DZP}. A Brillouin zone sampling of 24
inequivalent $k$ points, and a real-space grid equivalent to a plane-wave 
cutoff of 225 Ry (up to 82 $k$ points and 300 Ry in the convergence tests) have
been employed. This guarantees the convergence of the total energy within 
$\sim 0.1$ meV per atom in the supercell.

The Si(335)-Au system has been modeled by slabs containing up to four silicon 
double layers plus reconstructed surface layer. All the atomic positions were
relaxed except the bottom layer. The Si atoms in the bottom layer were 
saturated with hydrogen and remained at the bulk ideal positions during the
relaxation process. To avoid artificial stresses, the lattice constant of Si 
was fixed at the calculated bulk value, 5.42 \AA, which is very close to the
experimental value of 5.43 \AA.


\section{\label{structure} Structural model}

The total energy calculations show that it is energetically very favorable for 
the Au atoms to substitute into the top Si layer. The surface energy gain per unit 
cell is more than 1 eV as compared to adsorption above the surface. Furthermore, 
the Au substitution in the terraces is more stable than adsorption of the Au atoms 
at the step edge (by about $0.5$ eV per unit cell). Similar conclusions have been 
obtained for the case of Si(557)-Au surface, where the Au atoms prefer to 
substitute in the topmost Si layer, too \cite{Crain_2,Portal_1,Portal_2,Portal_3}. 
Therefore, in the following, I will focus on the structural models featuring the 
Si top layer atoms substituted by the gold. 

\begin{center}
\begin{figure}[h]
 \resizebox{0.9\linewidth}{!}{
  \includegraphics{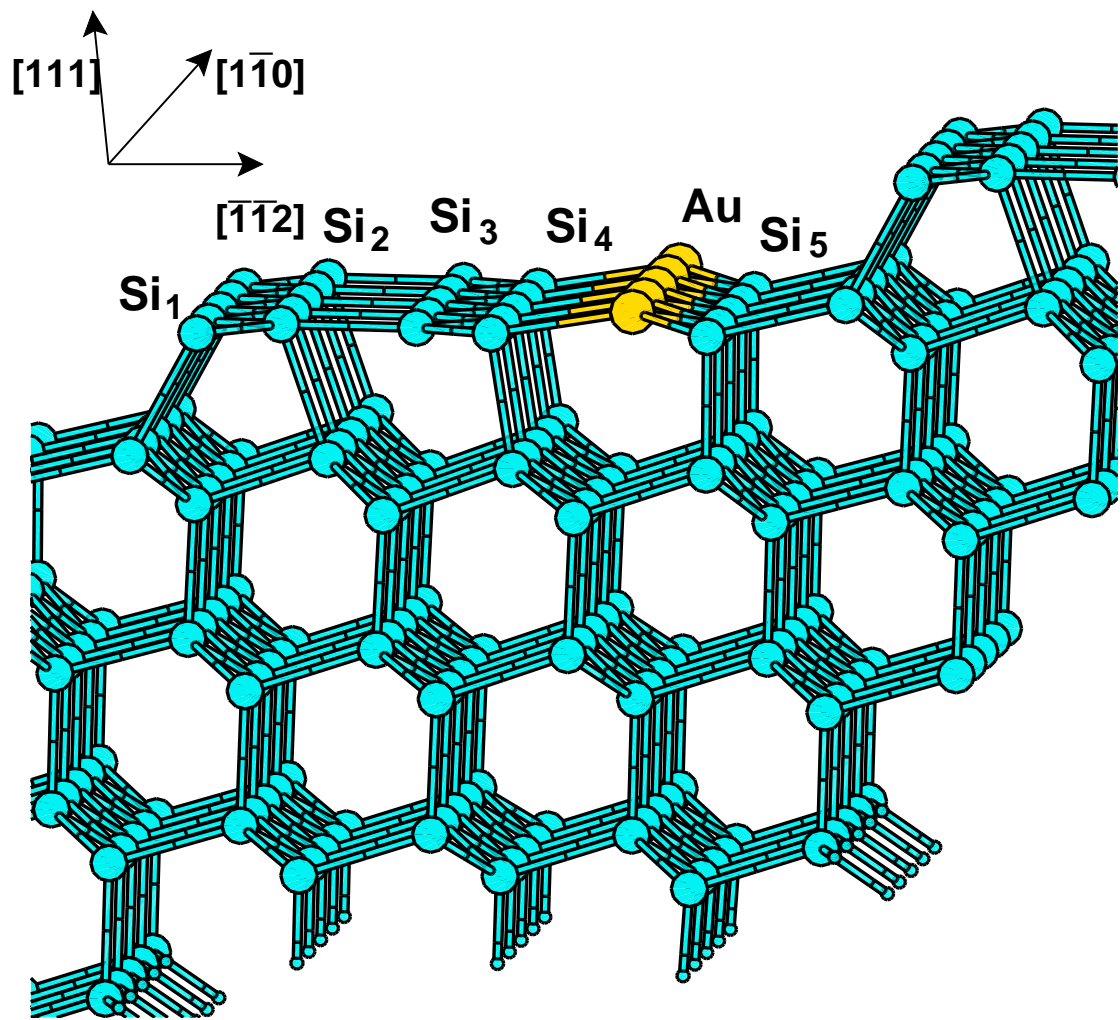}
}
 \caption{\label{Fig1} The most stable model of the Si(335)-Au surface. See the
 text for details.}
\end{figure}
\end{center}

The most stable model is shown in Fig. \ref{Fig1}. However, the other models, 
in which the Au atoms occupy various top layer silicon positions, from Si$_1$ 
to Si$_5$ (see Fig. \ref{Fig1} for labeling), have comparable energies. The 
differences are usually less than $\sim 0.7$ eV per unit cell, and the next 
'best' structural model, in which the gold occupies the Si$_3$ position, 
differs in energy by 182 meV only. The total energies of the above models with 
respect to the most stable model are summarized in Table \ref{Tab1}.

The present model is simply the model proposed by Crain {\it et al.} 
\cite{Crain_2}, which has not been deduced from any total energy calculations 
but was based on an analogy with Si(557)-Au model\,---\,a simple truncation of 
Si(557)-Au surface. Here, the DFT calculations confirm that this model is a 
good candidate for Si(335)-Au surface reconstruction. As one can read off from 
Fig. \ref{Fig1}, the Au atoms sit in the middle of the terrace and the rigidity of 
the Si structure keeps the Au wire stable against dimerization. The calculated 
Au-Si$_4$ bond length 2.43 {\AA} and 2.38 {\AA} for Au-Si$_5$ (see Fig. \ref{Fig1}) 
is quite close to calculated bulk Si-Si distance of 2.35 \AA, indicating that 
Au atoms affect the Si structure very little. Another important feature is a strong 
rebonding of the Si atoms near the step edge. The step edge atoms tend to 
saturate the dangling bonds in the neighboring terrace. As a result, a sort of 
'honeycomb' building block is created at the step edge, originally proposed for 
the alkali-induced $3 \times 1$ reconstruction of Si(111) surface 
\cite{Erwin_2}. It turns out that this sub-structure is a common feature of 
other Au-decorated Si vicinal surfaces \cite{Crain_2}.


\section{\label{band} Band structure}

\begin{center}
\begin{figure}
 \resizebox{\linewidth}{!}{
  \includegraphics{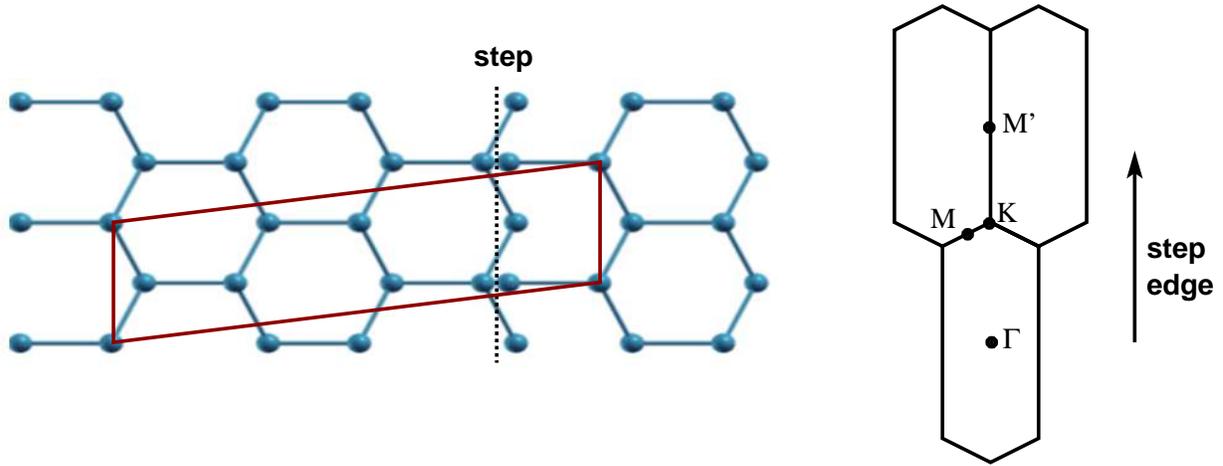}
}
 \caption{\label{Fig2} Surface unit cell (left panel) and the corresponding
 Brillouin zone (right panel) of the Si(335)-Au surface.}
\end{figure}
\end{center}

\begin{center}
\begin{figure}
 \resizebox{\linewidth}{!}{
  \includegraphics{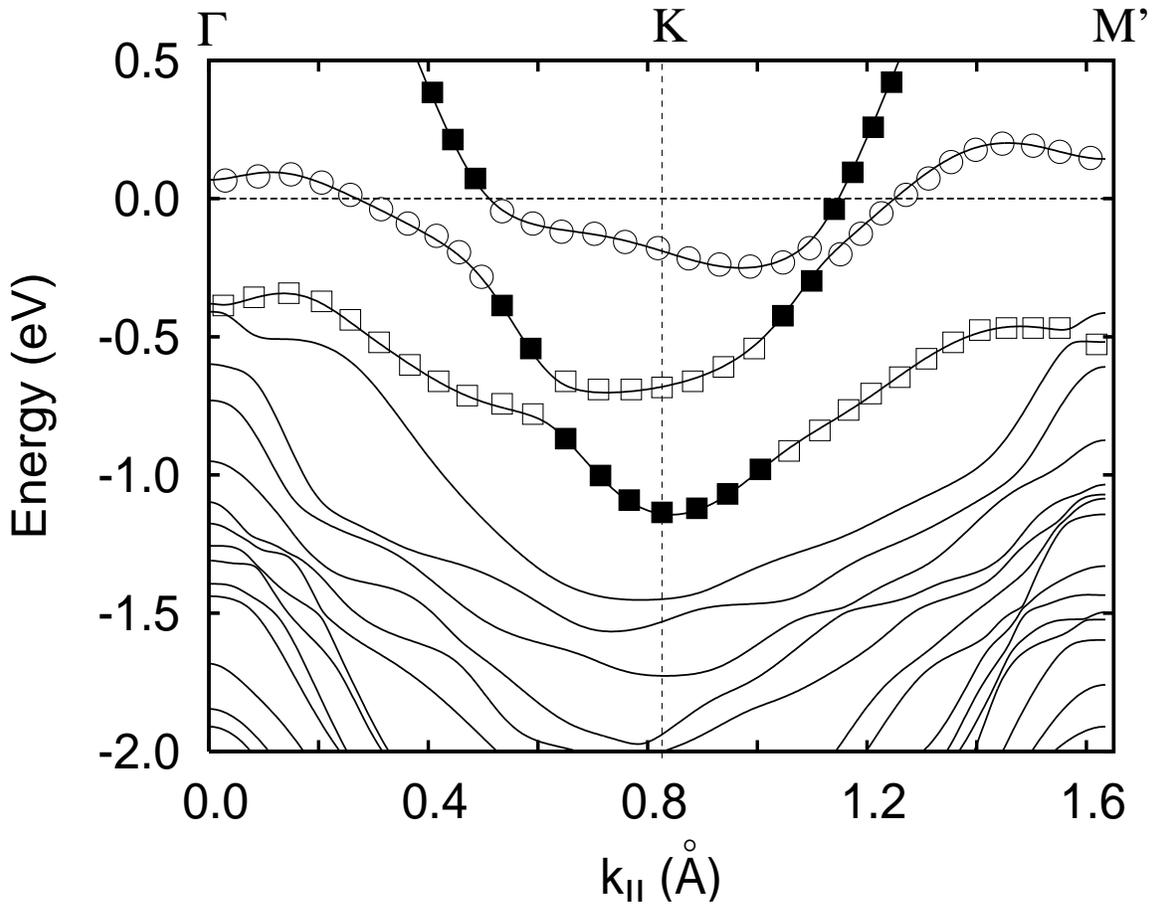}
}
 \caption{\label{Fig3} Calculated band structure along high symmetry lines in
         the two-dimensional Brillouin zone ($\Gamma$--K--M'). The band marked 
	  with open circles comes from the step edge Si atoms, while those 
	  marked with squares\,---\,from the Au and the neighboring Si atoms in the 
	  middle of the terrace.}
\end{figure}
\end{center}

The calculated band structure for this structural model, along the high 
symmetry line of two-dimensional Brillouin zone (Fig. \ref{Fig2}), is shown in 
Fig. \ref{Fig3}. The line defined by points $\Gamma$, K and M' is parallel to 
the steps, i.e. it goes along the Au chains. The Mulliken population analysis
\cite{Mulliken} has been performed in order to identify the main character of
the bands. Although this analysis is not completely unambiguous, it is
particularly useful for surface states. The band marked with open circles in
Fig. \ref{Fig3}, pinning the Fermi energy, comes from unsaturated bonds of the 
Si atoms at the step edge. A rather flat band around 0.5 eV below the Fermi 
energy (open squares), as well as a more dispersive band crossing the Fermi 
energy (filled squares) have the Au character. This picture looks a little bit 
counterintuitive, as one would expect a single band associated with $6s$
electrons of gold. In fact, these bands come from the Si atoms neighboring to 
the Au chain (Si$_4$ and Si$_5$ in Fig. \ref{Fig1}), and have gold character 
due to the hybridization. The 6s state of gold is well below the Fermi energy, 
thus completely occupied. The flat band marked with open squares comes from 
the Si$_4$ atoms, while the other, more dispersive one (filled squares), comes 
form the Si$_5$ atoms. There is one more band worth mentioning, not shown in 
Fig. \ref{Fig3}, having its origin in unsaturated bonds of the Si$_3$ atoms.
This band is also very flat, indicating its surface nature, and is located 
around 1 eV above the Fermi level. All the bands discussed above have also been 
identified in the case of Si(557)-Au surface \cite{Portal_1,Portal_2}. This 
similarity can be easily understood if one recalls that the Si(335)-Au surface 
is a truncation of the Si(557)-Au surface, as discussed previously. Thus, 
similar bands, although having different $k_{\parallel}$ dependence, should 
also be observed here.

\begin{center}
\begin{figure}
 \resizebox{\linewidth}{!}{
  \includegraphics{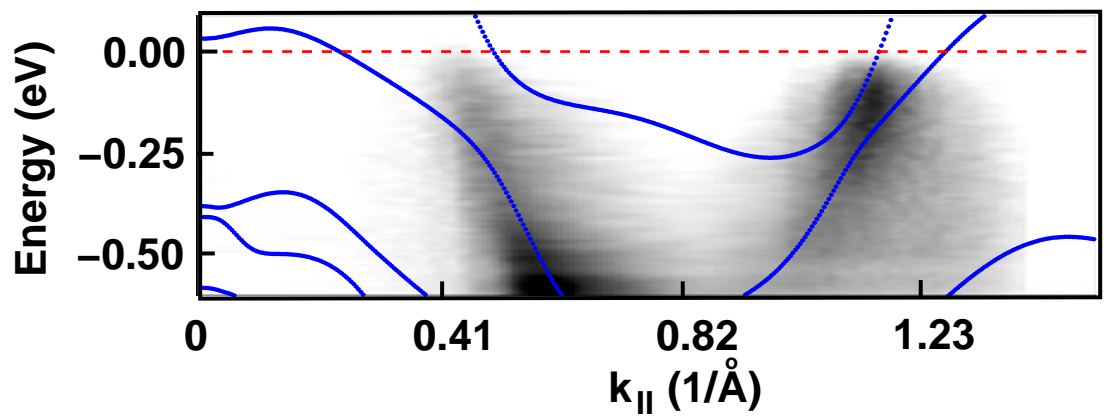}
}
 \caption{\label{Fig4} A comparison of the measured photoemission intensity along 
         the  $\Gamma$--K--M' line (Ref. \cite{Crain_2}) and the calculated band 
	  structure within the present model. Note that $\Gamma$ point
	  corresponds to $k_{\parallel} = 0$, K to $k_{\parallel} = 0.82$
	  \AA$^{-1}$, and M' to $k_{\parallel} = 1.64$ \AA$^{-1}$. The 
	  photoemission intensity is plotted  in a gray scale with high 
	  intensity shown dark.}
\end{figure}
\end{center}

A comparison of the calculated band structure with the photoemission spectra 
of Ref. \cite{Crain_2} is shown in Fig. \ref{Fig4}. As one can see, the present DFT 
calculations agree very well with the experimental data. In particular, there 
are two bands crossing the Fermi energy, one associated with the step edge Si
atoms, and the other one coming mainly from the Si atoms neighboring to the 
Au chain. Moreover, the shape of those bands and the values of $k_{\parallel}$, 
for which the bands cross the Fermi level, remain in good agreement with
experiment. Unfortunately, a more detailed comparison between present 
calculations and experiment of Ref. \cite{Crain_2} is not possible, as the 
photoemission spectra were measured in the energy window between $E_F$ and -0.5 
eV. However, I expect the agreement not be worse for lower energies.

\begin{center}
\begin{figure}
 \resizebox{\linewidth}{!}{
  \includegraphics{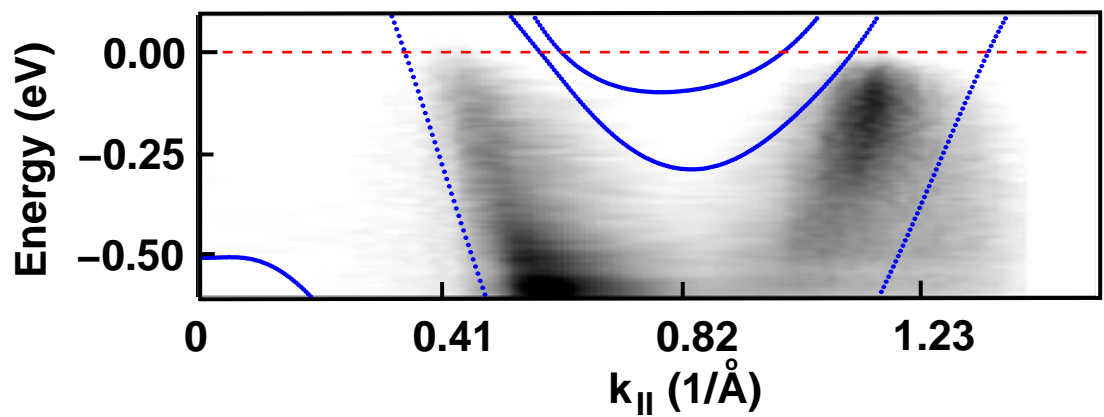}
}
 \caption{\label{Fig5} A comparison of the measured photoemission intensity along 
          the $\Gamma$--K--M' line (Ref. \cite{Crain_2}) and the calculated band 
	  structure within the model in which the top layer Si$_3$ atoms are
	  substituted by the gold atoms.}
\end{figure}
\end{center}

At this point I would like to comment on the band structure of the other models
studied here. Since the energy differences between all these models are rather 
small, it is natural to compare the band structure with the ARPES data. This 
would be the convincing criterion that the structural model is the correct one 
or at least is very close to the true Si(335)-Au surface reconstruction. Figure 
\ref{Fig5} shows a comparison of the ARPES data (Ref. \cite{Crain_2}) and the 
band structure calculated for the next 'best' structural model, in which the Au 
atoms occupy the Si$_3$ positions (see Fig. \ref{Fig1}). This model does not 
give as good agreement with the experimental data as the original one. In 
particular, there are three electronic bands crossing the Fermi energy, and 
none of them crosses the $E_F$ at the correct $k_{\parallel}$. The other models 
studied here also give wrong values of $k_{\parallel}$ at which the electronic 
bands cross the Fermi energy. Thus, this is an additional argument supporting 
the validity of the model shown in Fig. \ref{Fig1}.


\section{\label{conclusions} Conclusions}

In conclusion, the structural and electronic properties of Si(335)-Au surface
have been discussed within the density functional theory. The DFT calculations
revealed that the most stable structural model contains one Au atom per unit
cell, which substitutes the Si atom in the middle of terrace in the surface 
layer. The calculated electronic structure agrees well with photoemission
experimental data, showing two metallic bands. The less dispersive band 
comes from the Si atoms at the step edge, while the other one originates from 
hybridization between the Au and the neighboring Si atoms in the middle of the 
terrace in the surface layer.


\section*{Acknowledgements}
I would like to thank Prof. M. Ja\l ochowski for valuable discussions and 
critical reading of the manuscript. This work has been supported by the Polish 
Ministry of Education and Science under Grant No. N202 081 31/0372.



\newpage

\section*{Tables}

\begin{center}
\begin{table}[h]
\caption{\label{Tab1} Total energies of the structures of the Si(335)-Au surface
         for various positions of the Au atoms, as labeled in Fig. \ref{Fig1}. 
	 The energies (in eV per unit cell) are relative to the most stable 
	 model, shown in Fig. \ref{Fig1}.}
\begin{center}
\begin{tabular}{cccccc}
\hline
\hline
& & & & & \\
Au substitution: & Si$_1$ & Si$_2$ & Si$_3$ & Si$_4$  & Si$_5$\\

& & & & & \\
total energy: & 0.337 & 0.608 & 0.182 & 0.384 & 0.668 \\

& & & & & \\
\hline
\hline
\end{tabular}
\end{center}
\end{table}
\end{center}
%

\newpage

\section*{Figures}


\end{document}